\documentclass[11pt,a4,authoryear]{article}
\usepackage{amssymb, amsmath}
\usepackage{graphicx}
\usepackage{amsthm}
\usepackage{makeidx}
\usepackage{caption}
\usepackage{latexsym}
\usepackage{indentfirst}
\usepackage[T1]{fontenc}
\usepackage[applemac]{inputenc}
\usepackage{mathrsfs}
\usepackage{indentfirst}
\usepackage{amssymb,amsmath}
\usepackage{amsthm}
\usepackage{makeidx}
\usepackage{array}
\usepackage{graphicx}
\usepackage{latexsym}
\usepackage{caption}
\usepackage{fancyhdr}
\usepackage{bm}

\begin{document}

\newtheorem{Def}{Definition}
\newtheorem{Remark}[Def]{Remark}
\newtheorem{Ex}[Def]{Example}
\newtheorem{Theorem}[Def]{Theorem}
\newtheorem{Lemma}[Def]{Lemma}
\newtheorem{Corollary}[Def]{Corollary}
\newtheorem{Proposition}[Def]{Proposition}
\newtheorem{Figure}[Def]{Figure}
\newenvironment{Rem}{\begin{Remark}\rm}{\end{Remark}}
\newenvironment{Example}{\begin{Ex}\rm}{\end{Ex}}
\newenvironment{Proof}{\noindent{\bf  Proof.}}{\QED\bigskip}
\renewcommand{\theequation}{\thesection.\arabic{equation}}
\renewcommand{\theDef}{\thesection.\arabic{Def}}
\newcommand{\Enn}{{\rm I\! N}}
\newcommand{\erre}{{\rm I\! R}}
\newcommand{\Zeta}{{\sf Z\!\!Z}}
\newcommand{\QED}{\hfill $\Box$}
\newcommand{\Real}{\mathbb{R}}
\newcommand{\Exp}{\mathbb{E}}
\newcommand{\Prob}{\mathbb{P}}

\begin{center}
\begin{large}
\textbf{A Semi-Markov Modulated Interest Rate Model}
\end{large}
\end{center}

\begin{center}
{\sc Guglielmo D'Amico}\\
Department of Pharmacy, University "G. D'Annunzio",\\
via dei Vestini, 31, 66013 Chieti, Italy\\
e-mail: \texttt{g.damico@unich.it}\\
{\sc Raimondo Manca}\\
MEMOTEF Department, University "La Sapienza",\\
via del Castro Laurenziano, 9, 00000 Roma, Italy\\
e-mail: \texttt{raimondo.manca@uniroma1.it}\\
{\sc Giovanni Salvi}\\
MEMOTEF Department, University "La Sapienza",\\
via del Castro Laurenziano, 9, 00000 Roma, Italy\\
e-mail: \texttt{giovanni.salvi@uniroma1.it}\\
\end{center}

\indent {\bf Abstract:} In this paper we propose a semi-Markov modulated model of interest rates. We assume that the
switching process is a semi-Markov process with finite state space $E$ and the modulated process is a diffusive process.
We derive recursive equations for the higher order moments of the discount factor and we describe a Monte Carlo algorithm to execute simulations. The results are specialized to classical models as those by Vasicek, Hull and White and CIR with a semi-Markov modulation.\\ 


\indent {\bf Mathematical Subject Classification 2000:} .

\setcounter{equation}{0} \setcounter{Def}{0}
\section{Introduction.}
\label{Pref} \setcounter{equation}{0} \setcounter{Def}{0}

\noindent The literature on interest rate models is ample and mainly concerns models based on short rate dynamics and models of forward rate, see e.g. \cite{bjor04}. The advantages and the drawbacks of short rate models are well known in literature and even nowadays they receive continue attention by researcher and practitioners.\\
\indent The fundamental paper by \cite{vasi77} is the pioneering contribution of short rate models. Different papers presented extensions and alternatives to the Vasicek model. One interesting approach in term structure modelling is that of \cite{duff96} where it is assumed that the short rate process is a function of a state process. Interesting developments were subsequently obtained by \cite{mamo02} who characterized the term structure of a Markov interest rate model when the interest rate process is assumed to be a function of a continuous time non-homogeneous Markov chain. By using the forward measure in \cite{mamo04} it was shown how to price term structure derivative products.\\
\indent The present paper adopts the assumption that the short rate process is a diffusive process modulated by a continuous time semi-Markov process. The process of interest rates is defined by using the theory of semi-Markov reward processes with initial backward times as developed by \cite{sten07} and here opportunely extended to consider the case of stochastic permanence rewards. In this way we provide a general model of evolution which is able to reproduce a great variety of evolutions of the interest rate. The results include renewal type equations for the higher order moments of the zero coupon bond process and for the covariance function of the force of interest. Notice that, in the paper by \cite{hunt11}, a discrete time regime switching binomial-like model of the term structure, where the regime switches are governed by a discrete time semi-Markov process, is presented. There, it is possible to find a clear description of the advantages of the semi-Markov approach.\\
\indent It should be highlighted that our paper doesn't discuss the problem of transformation between the real-world (physical) measure and the pricing measure and for this reason the proposed model could be adopted not for pricing interest derivative products but for all applications that require the real-world evolution of rates, see \cite{rebo05} for a complete list. Finally notice that particular cases of our model were shown particularly adapted to solve problems with applications to insurance, see \cite{norb95}.\\
\indent The paper is organized as follows. The next section present a short description of semi-Markov processes with the introduction of the notation adopted in the analysis.  Section 3 presents the stochastic models of the short interest rates. In this section  we derive the main results concerning the equations for the higher order moments of the zero coupon bond process and for the covariance function of the force of interest. Section 4 describes the results specialized to popular diffusive models of short rates and present a Monte Carlo algorithm able to generate the synthetic data of the model.

Let us consider two sequences of random variables defined on a complete, filtered probability space $(\Omega, \mathcal{F},\mathcal{F}_{t}, P)$:\\
\indent (i)  $J_{n}:\Omega \rightarrow I=\{1,2,...,m\}$, $n\in {\rm \bf N}$ representing the state at the n-th transition;\\
\indent (ii) $T_{n}:\Omega \rightarrow {\rm \bf R}^{+}$ representing the time of the n-th transition.\\
\indent We suppose that $(J_{n},T_{n})$ is a Markov renewal process of kernel ${\rm \bf Q}=[Q_{ij}(t)]$:
\begin{equation*}
\begin{aligned}
& \Prob[J_{n+1}=j,T_{n+1}-T_{n}\leq t |\sigma(J_{a},T_{a}), J_{n}=i, 0\leq a \leq n] \\ 
& =\Prob \{J_{n+1}=j,T_{n+1}-T_{n}\leq t|J_{n}=i\}=: Q_{ij}(t).
\end{aligned}
\end{equation*}
We also know that:
\begin{displaymath}
p_{ij}:= \Prob \{J_{n+1}=j|J_{n}=i\}=\lim_{t\rightarrow \infty}Q_{ij}(t).
\end{displaymath}
\noindent ${\rm \bf P}=[p_{ij}]$ is the transition matrix of the embedded Markov chain $\{J_{n}\}_{n\in \Enn}$.\\
We assume that $\ t \rightarrow \mathbf{Q}(t),\  t \in \Real_{+}$ is absolutely continuous with respect to the Lebesgue measure on $\Real_{+}$. \\
\indent Denote the probability the process will leave state $i$ within time $t$ by:
\begin{displaymath}
H_{i}(t):= \Prob \{T_{n+1}-T_{n}\leq t|J_{n}=i\}=\sum_{k\in I}Q_{ik}(t).
\end{displaymath}
\indent It is possible to define the distribution function of the waiting time in each state $i$,
given that the state successively occupied is known:
\begin{eqnarray}
\label{condg}
G_{ij}(t)= \Prob \{T_{n+1}-T_{n}\leq t|J_{n}=i,J_{n+1}=j\}
\end{eqnarray}
\indent The related probabilities can be obtained by means of the following formula:
\begin{displaymath}
G_{ij} (t) = \left\{ {\begin{array}{ll}
   Q_{ij} (t) / p_{ij},\,\, if\,\,p_{ij}  \ne 0,  \\
   1,\,\, if\,\,\,\,p_{ij}  = 0, 
 \end{array} } \right.
\end{displaymath}
\indent Denote by $N(t):= \sup\{n:T_{n}\leq t\}$, the semi-Markov process $Z=(Z(t),\,t\in {\rm \bf R}_{+})$ is defined as $Z(t):=J_{N(t)}$.\\
\indent The transition probabilities of $Z(t)$ are defined in the following way:
\begin{equation}
\label{deftrprob}
\phi_{ij}(t):= \Prob \{Z(t)=j|Z(0)=i,T_{N(0)}=0\}.
\end{equation}
\indent They are obtained by solving the evolution equation:
\begin{equation}
\label{evoleqn}
\phi_{ij} (t) = \delta _{ij} (1 - H_{i}(t)) + \sum_{k\in I}\int_{0}^{t}\dot Q_{ik}(\theta)\phi _{kj} (t-\theta)d\theta
\end{equation}
\noindent where $\delta_{ij}$ represents the Kronecker symbol.\\
\indent The part $\delta _{ij} (1 - H_{i}(t))$ gives the probability that the system does not have transitions up to time $t$ given that it starts in state $i$ at time $0$.\\
\indent The term $\sum_{k\in I}\int_{0}^{t}\dot Q_{ik}(\theta)\phi _{kj} (t-\theta)d\theta $ considers the permanence of the system in state $i$ up to the time $\theta$ where a transition in state $k$ occurs. After the transition, the system will move to state $j$ following one of all the possible trajectories going from state $k$ to state $j$ in the remaining time $t-\theta$. All possible states $k$ and times $\theta$ are considered by the summation and the integration.\\
\indent Semi-Markov processes are very convenient to describe phenomena which display a duration effect. The duration effect affirms that the time system spent in a state influence its transition probabilities. One way to detect and quantify this effect with semi-Markov processes is by using backward and forward recurrence time processes. Recurrence processes were analyzed in \cite{jans07} and more recently in \cite{dami09} and \cite{dami11}.\\
\indent For our purposes we consider only the effect of the initial backward value.\\
\indent Given $(J_{n},T_{n})$, we define the following recurrence process:
\begin{eqnarray}
\label{backforw}
B(t):=t-T_{N(t)}.
\end{eqnarray}
\indent $B(t)$ is called the {\em backward time} (or age) process.\\
\indent The transition probabilities of a semi-Markov process change in function of the values of the backward time. In fact the conditional waiting times distribution functions $(\ref{condg})$ can be of any type and then, also no memoryless distributions can be used. In this case the time length spent in the starting state (initial backward value) changes the transition probabilities.\\
\indent The probability with starting backward $^{b}\phi_{ij}(u;t):=P[Z(t)=j|Z(0)=i,B(0)=u]$ denotes the probability of being in state $j$ after $t$ periods given that at present the process is in state $i$ and it entered into this state with the last transition $u$ periods before. It results that
\begin{equation}
\label{backin}
^{b}\phi_{ij}(u;t) =  \delta_{ij}\frac{1-H_{i}(u+t)}{1-H_{i}(u)}+
\sum_{k\in I}
\int_{0}^{t}\frac{\dot{Q}_{ik}(u+\theta)}{1-H_{i}(u)}\phi_{kj}(t-\theta)d\theta.
\end{equation}
\indent If $B(0)=0$, then $u=0$ and equation $(\ref{backin})$ degenerates in $(\ref{evoleqn})$.\\ 
\indent The probability with starting backward is a very special case of the transition probabilities studied in \cite{dami09} and \cite{dami11}.

\setcounter{equation}{0} \setcounter{Def}{0}
\section{The semi-Markov modulated model}
\label{Pref3} \setcounter{equation}{0} \setcounter{Def}{0}

In this section we define a semi-Markov modulated model of interest rates and we assess its probabilistic behaviour.\\
\indent In the following we assume that the force of interest at any time $t$ is a stochastic process of diffusive type whose parameters depend on the state of the semi-Markov process, on the backward recurrence time process and on the initial value of the force of interest $r_{0}$. To be more precise we assume that within two transition times $T_{n-1}$ and $T_{n}$ of the semi-Markov process the evolution of the force of interest follows the dynamic 
of a diffusive process whose parameters depends on the state $J_{n-1}$ of the semi-Markov process. 
The dynamic of the force of interest, between two consecutive renewal moments, will be indicated as
\begin{eqnarray}
\label{diffusive}
dr(t)=b_{i}(r(t),t)dt+\sigma_{i}(r(t),t) dW(t),\,\,\,r(0)=r_{0}
\end{eqnarray}
where it is supposed that $T_{0} = 0$, $T_{1} > t$ and $J_{0} = i$. The process $W(t)$ is a brownian motion with respect its own filtration $\mathcal{F}^{W}_{t}$, and we denote by $b$ and $\sigma$ the drift and the diffusion coefficients, respectively. 
The solution of the stochastic differential equation (\ref{diffusive}) will be denoted by $r_{i,r_{0}}(s)$, for $s \in [T_{0},T_{1})$. In $T_{1}$ the semi-Markov process transits to another state, say $j$, the force of interest in the time interval between $T_{1}$ and the next transition will evolve according to
\begin{eqnarray}
dr(t)=b_{j}(r(t),t)dt+\sigma_{j}(r(t),t) dW(t),\,\,\,r(T_{1})=r_{i,r_{0}}(T_{1})
\end{eqnarray}
for $t \in [T_{1},T_{2})$, and so on, then the resulting force of interest will be a continuos process. \\
We would like to stress that for any $s \in [T_{0},T_{1})$, and in general between any couple of transition times, the solution $r_{i,r_{0}}(s)$ is obtained with the standard methods of stochastic calculus.


\indent Due to the fact that the force of interest depends on the modulating process, we need to describe the force of interest process at any time $s$ given the information available at the present time (time zero as long as we work with an homogenous time model) given by the triplet of values $\{Z(0)=i,B(0)=u,r(0)=r_{0}\}$. To this end we can define the force of interest at the generic time $s$ as follows:
\begin{equation}
\begin{aligned}
\label{force}
&\delta_{i,u,r_{0}}(s) \stackrel{d}{=}\chi(T_{1}>s|J_{0}=i,T_{0}=-u,T_{1}>0)r_{i,r_{0}}(s) \\ 
&+ \chi(T_{1}\leq s|J_{0}=i,T_{0}=-u,T_{1}>0)\delta_{J_{1},0,r_{i,r_{0}}(T_{1})}(s-T_{1}) .
\end{aligned}
\end{equation}
Here $\chi ( A \ | \ B)$ is the indicator function of set $A$ given the information $B$. 

\begin{Remark}
\label{rem1}
The processes $\chi (T_{1}>s|J_{0}=i,T_{0}=-u,T_{1}>0)$ and $r_{i,r_{0}}(s)$ are independent for any $s \in \Real_{+}$ and $i \in I$.
\end{Remark}

\begin{Remark}
\label{rem2}
The process $\chi (s < T_{1} < s +h | J_{0}=i,T_{0}=-u,T_{1}>0) \delta_{J_{1},0,r_{i,r_{0}}(T_{1})}(s+h-T_{1})$ conditioning to the value of $r_{i,r_{0}}(T_{1})$ is independent of $r_{i,r_{0}}(s)$.
\end{Remark}

\indent We are interested in the ZCB process $v_{i,u,r_{t}}(t,T):=exp(-\int_{t}^{T}\delta_{i,u,r_{t}}(s)ds)$ expressing the value of a zero coupon bond paying $1$ Euro at time $T$ given that at current time $t$ the semi-Markov process is in state $i$ and it is entered in this state $u$ periods before and the force of interest at that time is $r_{t}$. To this end in what follow we would like to consider only diffusion processes that allow an explicit representation of the Laplace transform of $\int_{0}^{t}r_{i,r_{0}}(s)ds$ for example \cite{vasi77}, \cite{hull} and Cox, Ingersoll and Ross (1985) (CIR) model.

\begin{Theorem}
\label{teorema1}
Let 
\begin{equation}
V_{i,u,r_{t}}^{(n)}(t,T)= \Exp [(v_{i,u,r_{t}}(t,T))^{n}]
\end{equation} 
be the nth order moment of the ZCB process 
\noindent then it results that
\begin{equation}
\begin{aligned}
& V_{i,u,r_{t}}^{(n)}(t,T)=\frac{1-H_{i}(T-t+u)}{1-H_{i}(u)}\big(B_{i,r_{t}}^{(n)}(t,T)\big) \\ 
&+ \sum_{k\in I}\int_{t}^{T}\frac{\dot{Q}_{ik}(\tau + u)}{1-H_{i}(u)}\bigg(B_{i,r_{t}}^{(n)}(t,\tau)\int_{-\infty}^{+\infty}V_{k,0,x}^{(n)}(\tau,T)F_{r_{i,r_{t}}(\tau)}(dx)\bigg)d\tau ,
\end{aligned}
\end{equation}
where $B_{i,r_{t}}^{(n)}$ is defined by
\begin{equation}
B_{i,r_{t}}^{(n)} (t,T) \doteq \Exp \left \{ exp\left( -n \int_{t}^{T}r_{i,r_{t}}(s)ds \right) \right\}
\end{equation} 
\end{Theorem}
\noindent {\bf Proof.}
Let consider, at the time $t$, the process $v_{i,u,r_{t}}(t,T):=exp(-\int_{t}^{T}\delta_{i,u,r_{t}}(s)ds)$ and condition to the time of next transition of the switching process. We can partition the state space in two possible events $\{ T_{N(t)+1}>T \}$ or $\{ T_{N(t)+1}\leq T \}$.\\
The first event $\{ T_{N(t)+1}>T \}$ corresponds to the possibility to have no transition up to the time $T$ and it has probability $\frac{1-H_{i}(T-t+u)}{1-H_{i}(u)}$. Under this event, the force of interest is given by
\begin{equation}
\delta_{i,u,r_{t}}(s)=r_{i,r_{t}}(s) \qquad \qquad \textrm{for} \ s \in [t,T] .
\end{equation}
In this case, the ZCB process assumes value
\begin{equation}
v_{i,u,r_{t}}(t,T) \bigg |_{T_{N(t)+1}>T} = exp \left( -\int_{t}^{T}r_{i,r_{t}}(s)ds \right) ,
\end{equation}
tacking now the expectation we get
\begin{equation}
\Exp \{ (v_{i,u,r_{t}}(t,T))^{n} | T_{N(t)+1} > T \} =  B_{i,r_{t}}^{(n)} (t,T) .
\end{equation}
\indent Then we have that 
\begin{eqnarray}
\label{pezzo1}
& & \Exp \{ (v_{i,u,r_{t}}(t,T))^{n}\chi (T_{N(t)+1}>T|J_{N(t)}=i,T_{N(t)}=t-u,T_{N(t)+1}>t, r(t)=r_{t}) \} \nonumber \\
& = &  \frac{1-H_{i}(T-t+u)}{1-H_{i}(u)} B_{i,r_{t}}^{(n)} (t,T). 
\end{eqnarray}

The other event $\{ T_{N(t)+1}\in (t,T] , J_{N(t)+1}\in I \}$ corresponds to the possibility to have at least one transition in the considered time interval. The probability that the semi-Markov process has the first transition in the time interval $( \tau, \tau+d\tau )$, for $\tau \in (t,T)$, into a state $k \in I$, is given by
  \begin{eqnarray*}
 P(J_{N(t)+1}=k, T_{N(t)+1}\in (\tau, \tau+d\tau)|\mathcal{F}^{SM}_{t},Z(t)=i,B(t)=u)
  = \frac{\dot{Q}_{ik}(\tau + u)}{1-H_{i}(u)}d\tau ,
  \end{eqnarray*}
where $(\mathcal{F}^{SM}_{t})_{t}$ is the filtration generated by the process $(Z,B)$, that is the semi-Markov and the backward recurrence time together. To properly evaluate the expectation under this event, we have to consider even all the possible value that the force of interest can assume on transition time $\tau$. Denoting by $F_{r_{i,r_{t}}(\tau)}$ the cumulative distribution function associated to $r_{i,r_{t}} (\tau)$, using the continuity property of the force of interest, we have that
\begin{eqnarray*}
\hspace{-10pt}P(J_{N(t)+1}=k, T_{N(t)+1}\in (\tau, \tau+d\tau),\delta_{i,u,r_{t}}(\tau)\in (x,x+dx)|\mathcal{F}_{t})  = \frac{\dot{Q}_{ik}(\tau + u)}{1-H_{i}(u)}F_{r_{i,r_{t}}(\tau)}(dx)d\tau
\end{eqnarray*}
where $\mathcal{F}_{t}$ is the filtration generated by (Z,B,W). Under the event $A = \{J_{N(t)+1}=k, T_{N(t)+1}\in (\tau, \tau+d\tau),\delta_{i,u,r_{t}}(\tau)\in (x,x+dx), \mathcal{F}_{t} \}$ the ZCB process can be expressed as
\begin{equation*}
v_{i,u,r_{t}}(t,T) \bigg |_{A} =exp \left( -\int_{t}^{T}\delta_{i,u,r_{t}}(s)ds \right) \Bigg |_{A} = exp \left( -\int_{t}^{\tau}r_{i,r_{t}}(s)ds \right) exp \left( -\int_{\tau}^{T}\delta_{k,0,x}(s)ds \right) \Bigg |_{A} .
\end{equation*}
Moreover from remark \ref{rem2}, we have
\begin{eqnarray*}
& & \hspace{-20pt} \Exp \{ (v_{i,u,r_{t}}(t,T))^{n} | A \} = \Exp \left\{ exp\left( -n\int_{t}^{\tau}r_{i,r_{t}}(s)ds \right) exp \left( -n\int_{\tau}^{T}\delta_{k,0,x}(s)ds \right) | A \right\} \\
& = & \Exp \left\{ exp\left( -n\int_{t}^{\tau}r_{i,r_{t}}(s)ds \right) | A \right\} \Exp \left\{ exp \left( -n\int_{\tau}^{T}\delta_{k,0,x}(s)ds \right) | A \right\} = B_{i,r_{t}}^{(n)} (t,\tau) V_{k,0,x}^{(n)}(\tau,T) .
\end{eqnarray*}
and consequently 
\begin{equation}
\label{pezzo2}
\begin{aligned}
& \hspace{-10pt}\Exp \{ (v_{i,u,r_{t}}(t,T))^{n} \chi (t< T_{N(t)+1}\leq T|J_{N(t)}=i,T_{N(t)}=t-u,T_{N(t)+1}>t, r(t)=r_{t})\} \\
& = \sum_{k\in I}\int_{t}^{T}\frac{\dot{Q}_{ik}(\tau + u)}{1-H_{i}(u)}\bigg(B_{i,r_{t}}^{(n)}(t,\tau)\int_{-\infty}^{+\infty}V_{k,0,x}^{(n)}(\tau,T)F_{r_{i,r_{t}}(\tau)}(dx)\bigg)d\tau.
\end{aligned}
\end{equation}

\noindent The value of the zero coupon bond is given by
\begin{equation}
\label{zcb}
\begin{aligned}
 \Exp \{ (v_{i,u,r_{t}}(t,T))^{n} \} & =  \Exp \{ (v_{i,u,r_{t}}(t,T))^{n} \chi (T_{N(t)+1}> T|J_{N(t)}=i,T_{N(t)}=t-u,T_{N(t)+1}>t, r(t)=r_{t})\}\\
& +\Exp \{ (v_{i,u,r_{t}}(t,T))^{n} \chi (t< T_{N(t)+1}\leq T|J_{N(t)}=i,T_{N(t)}=t-u,T_{N(t)+1}>t, r(t)=r_{t})\} ,
\end{aligned}
\end{equation}
then by substitution of expressions $(\ref{pezzo1})$ and $(\ref{pezzo2})$ in $(\ref{zcb})$ the proof is complete.
\begin{flushright} $\Box$ \end{flushright}

\begin{Corollary}
\label{corollario1}
Let denote by $R_{i,u,r_{t}}(s)= \Exp [\delta_{i,u,r_{t}}(s)]$, then it results that
\begin{equation}
\begin{aligned}
 R_{i,u,r_{t}}(s)=\frac{1-H_{i}(s+u)}{1-H_{i}(u)}m_{i,r_{t}}(t,s) +\sum_{k\in I}\int_{0}^{s}\frac{\dot{Q}_{ik}(\tau + u)}{1-H_{i}(u)} \bigg( \int_{-\infty}^{+\infty}R_{k,0,x}(s-\tau) F_{r_{i},r_{t}(\tau)}(dx) \bigg)d\tau
\end{aligned}
\end{equation}
here, $m_{i,r_{t}}(s) := \Exp [ r_{i,r_{t}} (s) ]$.
\end{Corollary}
We can derive the following result that help us on obtaining the covariance function of the force of interest.
\begin{Theorem}
\label{teorema2}
Let $s>0$ and $h>0$, and denote by
\begin{equation}
\Xi_{i,u,r_{0}}(s,h)= \Exp [\delta(s)\delta(s+h)|Z(0)=i,B(0)=u,\delta(0)=r_{0}].
\end{equation} 
\indent Then it results that
\begin{equation}
\begin{aligned}
& \Xi_{i,u,r_{0}}(s,h)=\frac{1-H_{i}(s+h+u)}{1-H_{i}(u)}\rho_{i,r_{0}}(s,s+h)\\ 
&+\sum_{k\in I}\int_{s}^{s+h}\frac{\dot{Q}_{ik}(\tau +u)}{1-H_{i}(u)} m_{i,r_{0}} (s) \int_{-\infty}^{+\infty}
R_{k,0,x}(s+h- \tau) F_{i,r_{0}(\tau)}(dx) d\tau \\
& +\sum_{k\in I}\int_{0}^{s} \frac{\dot{Q}_{ik}(\tau +u)}{1-H_{i}(u)}\int_{-\infty}^{+\infty}\Xi_{k,0,x}(s-\tau, s+h-\tau) F_{i,r_{0}(\tau )}(dx) d\tau
\end{aligned}
\end{equation}
here, $\rho_{i,r_{0}}(s,s+h) := \Exp [r_{i,r_{0}}(s)r_{i,r_{0}}(s+h)]$.
\end{Theorem}
\noindent {\bf Proof.}
Using the definition of force of interest process $\delta$, cf. formula $(\ref{force})$, we have that
\begin{equation}
\begin{aligned}
\label{force2}
& \delta_{i,u,r_{0}}(s)\delta_{i,u,r_{0}}(s+h) \stackrel{d}{=}\{\chi(T_{1}>s|J_{0}=i,T_{0}=-u,T_{1}>0)r_{i,r_{0}}(s)\\
&+ \chi(T_{1}\leq s|J_{0}=i,T_{0}=-u,T_{1}>0)[\delta_{J_{1},0,r_{i,r_{0}}(T_{1})}(s-T_{1})]\} \\ 
& \times \big\{ \chi(T_{1}>s+h|J_{0}=i,T_{0}=-u,T_{1}>0)r_{i,r_{0}}(s+h)\\
&+ \chi(T_{1}\leq s+h|J_{0}=i,T_{0}=-u,T_{1}>0)[\delta_{J_{1},0,r_{i,r_{0}}(T_{1})}(s+h-T_{1})]\big\} ,\\
\end{aligned}
\end{equation}
solving the product we obtain three terms
\begin{equation}
\begin{aligned}
\label{force2bis}
& \delta_{i,u,r_{0}}(s)\delta_{i,u,r_{0}}(s+h) = \chi(T_{1}>s+h|J_{0}=i,T_{0}=-u,T_{1}>0)r_{i,r_{0}}(s)r_{i,r_{0}}(s+h)\\
& + \chi(s< T_{1}\leq s+h|J_{0}=i,T_{0}=-u,T_{1}>0)r_{i,r_{0}}(s)\delta_{J_{1},0,r_{i,r_{0}}(T_{1})}(s+h-T_{1})\\
& + \chi(T_{1}\leq s|J_{0}=i,T_{0}=-u,T_{1}>0)\delta_{J_{1},0,r_{i,r_{0}}(T_{1})}(s-T_{1})\delta_{J_{1},0,r_{i,r_{0}}(T_{1})}(s+h-T_{1}) ,
\end{aligned}
\end{equation}
The first corresponds to the possibility that semi-Markov process has no transition up to time $s+h$, the second to the case to have the first transition into the time interval $(s,s+h)$ and the third takes into account for the possibility to have the first transition before $s$. \\
\indent  Taking expectations of the first term on right hand side (r.h.s.) of (\ref{force2bis}) and nothing that (cf. remark \ref{rem1}) the random variable $\chi(T_{1}>s+h|J_{0}=i,T_{0}=-u,T_{1}>0)$ is independent both from $r_{i,r_{0}}(s)$ and $r_{i,r_{0}}(s+h)$ yields,
\begin{equation}
\label{pz1}
\begin{aligned}
& \Exp [\chi(T_{1}>s+h|J_{0}=i,T_{0}=-u,T_{1}>0)r_{i,r_{0}}(s)r_{i,r_{0}}(s+h)] \\
& =\Exp [\chi(T_{1}>s+h|J_{0}=i,T_{0}=-u,T_{1}>0)] \Exp [r_{i,r_{0}}(s)r_{i,r_{0}}(s+h)] \\ 
&=\frac{1-H_{i}(s+h+u)}{1-H_{i}(u)}\rho_{i,r_{0}}(s,s+h) .
\end{aligned}
\end{equation}

Regarding the second term on r.h.s. of (\ref{force2bis}), tacking the expectation and conditioning on the value of $r_{i,r_{0}}(T_{1})$, we can use the independence between $r_{i,r_{0}}(s)$ and $\chi(s< T_{1}\leq s+h|J_{0}=i,T_{0}=-u,T_{1}>0)\delta_{J_{1},0,r_{i,r_{0}}(T_{1})}(s+h-T_{1})$ (cf. remark \ref{rem2}) to obtain
\begin{eqnarray*}
\label{pz2}
 & \Exp \{ \Exp [\chi(s< T_{1}\leq s+h|J_{0}=i,T_{0}=-u,T_{1}>0)r_{i,r_{0}}(s)\delta_{J_{1},0,r_{i,r_{0}}(T_{1})}(s+h-T_{1}) | r_{i,r_{0}}(T_{1}) = x]  \} \\
 & = \Exp \{ \Exp [\chi(T_{1}>s+h|J_{0}=i,T_{0}=-u,T_{1}>0)\delta_{J_{1},0,r_{i,r_{0}}(T_{1})}(s+h-T_{1}) | r_{i,r_{0}}(T_{1}) = x ] \} \Exp [r_{i,r_{0}}(s)] \nonumber \\
 & = \sum_{k\in I}\int_{s}^{s+h}\frac{\dot{Q}_{ik}(\tau + u)}{1-H_{i}(u)} \Exp [r_{i,r_{0}}(s)] \int_{-\infty}^{+\infty} \Exp [\delta_{k,0,x}(s+h-\tau)] F_{i,r_{0}(\tau)}(dx) d\tau, \nonumber
\end{eqnarray*}
\noindent but 
\begin{equation}
E[r_{i,r_{0}}(s)]=m_{i,r_{0}}(s),
\end{equation}
and 
\begin{equation}
\begin{aligned}
& \Exp [\delta_{k,0,x}(s+h-\tau)] = R_{k,0,x}(s+h-\tau) ,
\end{aligned}
\end{equation}
then, by substitution, we get
\begin{eqnarray}
\label{pz2bis}
 & E[\chi(s< T_{1}\leq s+h|J_{0}=i,T_{0}=-u,T_{1}>0)r_{i,r_{0}}(s)\delta_{J_{1},0,r_{i,r_{0}}(T_{1})}(s+h-T_{1})] \nonumber \\
 & = \sum_{k\in I}\int_{s}^{s+h}\frac{\dot{Q}_{ik}(\tau +u)}{1-H_{i}(u)} m_{i,r_{0}}(s) \int_{-\infty}^{+\infty}
R_{k,0,x}(s+h- \tau) F_{i,r_{0}(\tau)}(dx) d\tau .
\end{eqnarray}
\indent Finally, tacking the expectation of the third term on l.h.s. of (\ref{force2bis}) we obtain
\begin{eqnarray*}
\label{pz3}
 & E[\chi(T_{1}\leq s|J_{0}=i,T_{0}=-u,T_{1}>0)\delta_{J_{1},0,r_{i,r_{0}}(T_{1})}(s-T_{1})\delta_{J_{1},0,r_{i,r_{0}}(T_{1})}(s+h-T_{1}) ] \\
 & = \sum_{k\in I}\int_{0}^{s}\frac{\dot{Q}_{ik}(\tau + u)}{1-H_{i}(u)} \Exp [\delta_{k,0,r_{i,r_{0}}(\tau)}(s-\tau)\delta_{k,0,r_{i,r_{0}}(\tau)}(s+h-\tau)]d\tau, \nonumber
\end{eqnarray*}
and by considering all possible values of $r_{i,r_{0}}(\tau)$ we get
\begin{eqnarray*}
\label{pz3bis}
 & E[\chi(T_{1}\leq s|J_{0}=i,T_{0}=-u,T_{1}>0)\delta_{J_{1},0,r_{i,r_{0}}(T_{1})}(s-T_{1})\delta_{J_{1},0,r_{i,r_{0}}(T_{1})}(s+h-T_{1}) ] \\
 & = \sum_{k\in I}\int_{0}^{s}\frac{\dot{Q}_{ik}(\tau + u)}{1-H_{i}(u)}\Big(\int_{-\infty}^{+\infty} \Exp [\delta_{k,0,x}(s-\tau)\delta_{k,0,x}(s+h-\tau)] F_{i,r_{0}(\tau)}(dx) \Big) d\tau, \nonumber
\end{eqnarray*}
but
\begin{eqnarray*}
\Exp [\delta_{k,0,x}(s-\tau)\delta_{k,0,x}(s+h-\tau)] = \Xi_{k,0,x}(s-\tau, s+h-\tau) ,
\end{eqnarray*}
then
\begin{eqnarray*}
\label{pz3final}
 & E[\chi(T_{1}\leq s|J_{0}=i,T_{0}=-u,T_{1}>0)\delta_{J_{1},0,r_{i,r_{0}}(T_{1})}(s-T_{1})\delta_{J_{1},0,r_{i,r_{0}}(T_{1})}(s+h-T_{1}) ] \\
 & = \sum_{k\in I}\int_{0}^{s}\frac{\dot{Q}_{ik}(\tau + u)}{1-H_{i}(u)}\Big(\int_{-\infty}^{+\infty} \Xi_{k,0,x}(s-\tau, s+h-\tau) F_{i,r_{0}(\tau)}(dx) \Big) d\tau, \nonumber
\end{eqnarray*}
\begin{flushright} $\Box$ \end{flushright}

section{Particular Models}
So far, we do not specified any specific dynamic driving the force of interest between two consecutive renewal moments. In this section we would like to consider some particular diffusion models for the force of interest as \cite{vasi77}, \cite{hull} and \cite{coxi85}.
For these models, as well known, it is possible to obtain an explicit representation of the Laplace transform of the integral of the force of interest $r$. We will see as our general results specializes in these particular cases. Without loss of generality, as long as we are working with an homogeneous time model, we can suppose that the present time is $t=0$, with $J_{0} = i$, then we will study the dynamic of the force of interest from $0$ up to a generic instant $t < T_{1}$.

\subsection{Vasicek modulated model}
We assume that the process $r_{i,r_{0}}(t)$, satisfies
\begin{displaymath}
\left\{ \begin{array}{l}
dr_{i,r_{0}}(t) = a_{i}(b_{i}-r_{i,r_{0}}(t))dt + \sigma_{i} dW_{t} \\
r_{i,r_{0}}(0) = r_{0}
\end{array} \right.
\end{displaymath}
where $a_{i}$, $b_{i}$ and $\sigma_{i}$, for fixed $i \in I$, are non negative constants. The solution of the previous stochastic differential equation is
\begin{eqnarray}
r_{i,r_{0}}(t) = b_{i} + (r_{0}-b_{i})e^{-a_{i}t} + \sigma_{i} e^{-a_{i}t} \int_{0}^{t} e^{a_{i}s} dW_{s} .
\end{eqnarray}
For any fixed $t$, $r_{i,r_{0}}(t)$ is normal distributed, $r_{i,r_{0}}(t) \sim \mathcal{N} ( m_{i,r_{0}}(t), \sigma^{2}_{i,r_{0}}(t))$ with mean
\begin{eqnarray}
\label{Eq:meanVas}
m_{i,r_{0}}(t) = \Exp [ r_{i,r_{0}}(t) ] = b_{i} + (r_{0}-b_{i})e^{-a_{i}t}
\end{eqnarray}
and variance
\begin{eqnarray}
\label{Eq:varVas}
\sigma^{2}_{i,r_{0}}(t) = \frac{\sigma_{i}^{2}}{2a_{i}} (1 - e^{-2a_{i}t}) .
\end{eqnarray}
 It follows that (see \cite{Lamberton})
$\int_{0}^{t} r(s) ds$ is a normal random variable.
Then we can completely characterize the distribution of $\int_{0}^{t} r_{i,r_{0}}(s) ds$ with its mean and variance.
The mean, using Fubini's Theorem, is given by
\begin{eqnarray}
 \Exp \left[ \int_{0}^{t} r_{i,r_{0}}(s) ds \right] = \int_{0}^{t} \Exp [ r_{i,r_{0}}(s) ] ds = \int_{0}^{t} m_{i,r_{0}}(s) ds =  b_{i}t + \frac{r_{0}-b_{i}}{a_{i}} (1 - e^{-a_{i}t}),
\end{eqnarray}
and the variance by
\begin{eqnarray}
Var \left[ \int_{0}^{t} r_{i,r_{0}}(s) ds \right] = \frac{\sigma_{i}^{2}t}{a_{i}^{2}} - \frac{\sigma_{i}^{2}}{a_{i}^{3}}(1 - e^{-a_{i}t}) - \frac{\sigma_{i}^{2}}{2a_{i}^{3}}(1 - e^{-a_{i}t})^{2} .
\end{eqnarray}
Now, we can give an analytical expression of the $n^{th}$ moment of a ZCB value between two consecutive renewal moment, $B_{i,r_{0}}^{(n)}(0,t)$ (cf. theorem \ref{teorema1}), defined by 
\begin{eqnarray}
B^{(n)}_{i,r_{0}}(0,t) = \Exp \left[ exp \{-n\int^{t}_{0} r_{i,r_{0}}(t)dt \} \right]
\end{eqnarray}
which can be seen as the Laplace transform of the Normal random variable $\int_{0}^{T}r_{i,r_{0}}(t)dt$ and then it can be expressed as
\begin{eqnarray}
B^{(n)}_{i,r_{0}}(0,t) = exp \left\{ - n\Exp \left[ \int^{t}_{0} r_{i,r_{0}}(t)dt \right] + \frac{n^{2}}{2}  Var \left[ \int_{0}^{t} r_{i,r_{0}}(t) dt \right] \right\} .
\end{eqnarray}  
Substituting the expression for mean and variance we get
\begin{equation}
\begin{aligned}
B^{(n)}_{i,r_{0}}(0,t) = exp \left\{ \left( \frac{\sigma_{i}^{2}n^{2}}{a_{i}^{2}} - n b_{i} \right) t - \left( \frac{\sigma_{i}^{2}n^{2}}{a_{i}^{3}} + \frac{n(r_{0}-b_{i})}{a_{i}} \right) \left( 1 - e^{-a_{i}t} \right) - \frac{\sigma_{i}^{2}n^{2}}{2a_{i}^{3}}(1 - e^{-a_{i}t})^{2} \right\} .
\end{aligned}
\end{equation}
Now, we are able to express all the moments of the ZCB in any time interval.
\begin{Corollary}
The nth order moment of the ZCB process described by a Vasicek model modulated by a semi-Markov process of kernel $\mathbf{Q}$ is 
\begin{equation}
\label{Eq:ZCBVas}
\begin{aligned}
& V_{i,u,r_{t}}^{(n)}(t,T)=\Exp \left\{ exp(-n\int_{t}^{T}\delta_{i,u,r_{t}}(s)ds) \right\} = \frac{1-H_{i}(T-t+u)}{1-H_{i}(u)}\big(B_{i,r_{t}}^{(n)}(t,T)\big)\\
& + \sum_{k\in I}\int_{t}^{T}\frac{\dot{Q}_{ik}(\tau + u)}{1-H_{i}(u)}\bigg(B_{i,r_{t}}^{(n)}(t,\tau)\int_{-\infty}^{+\infty}f_{\mathcal{N}(m_{i,r_{t}}(\tau -t),\sigma_{i,r_{t}}^{2}(\tau -t))}(x)V_{k,0,x}^{(n)}(\tau,T)dx\bigg)d\tau .
\end{aligned}
\end{equation}
where $B_{i,r_{t}}^{(n)}(t,T)$ is given in equation (\ref{Eq:ZCBVas}) and $f_{\mathcal{N}(m_{i,r_{t}}(\tau -t),\sigma_{i,r_{t}}^{2}(\tau -t))}$ is the distribution function of a normal with mean $m_{i,r_{t}}(\tau -t)$ and variance $\sigma_{i,r_{t}}^{2}(\tau-t)$ (cf. eq. (\ref{Eq:meanVas}) and (\ref{Eq:varVas}) respectively).
\end{Corollary}
\begin{Corollary}
The first moment of the force of interest of the Vasicek model modulated by a semi-Markov process of kernel $\mathbf{Q}$ is
\begin{equation}
\begin{aligned}
& R_{i,u,r_{t}}(s)=\Exp [\delta_{i,u,r_{t}}(s)]=\frac{1-H_{i}(s+u)}{1-H_{i}(u)}\big(b_{i}+(r_{t}-b_{i})e^{-a_{i}s}\big)\\
& +\sum_{k\in I}\int_{0}^{s}\frac{\dot{Q}_{ik}(\tau + u)}{1-H_{i}(u)}\int_{-\infty}^{+\infty}f_{\mathcal{N}(m_{i,r_{t}}(\tau),\sigma_{i,r_{t}}^{2}(\tau ))}(x)R_{k,0,x}(s-\tau)dx\bigg)d\tau .
\end{aligned}
\end{equation}
where $f_{\mathcal{N}(m_{i,r_{t}}(\tau),\sigma_{i,r_{t}}^{2}(\tau))}$ is the distribution function of a normal with mean $m_{i,r_{t}}(\tau)$ and variance $\sigma_{i,r_{t}}^{2}(\tau)$ (cf. eq. (\ref{Eq:meanVas}) and (\ref{Eq:varVas}) respectively) .
\end{Corollary}
\begin{Corollary}
The product moment of the force of interest of a Vasicek model modulated by a semi-Markov process of kernel $\mathbf{Q}$ is
\begin{equation}
\begin{aligned}
& \Xi_{i,u,r_{0}}(s,h)=\Exp [\delta(s)\delta(s+h)|Z(0)=i,B(0)=u,\delta(0)=r_{0}] = \frac{1-H_{i}(s+h+u)}{1-H_{i}(u)}\rho_{i,r_{0}}(s,s+h)\\
& +\sum_{k\in I}\int_{s}^{s+h}\frac{\dot{Q}_{ik}(\tau +u)}{1-H_{i}(u)}\big(b_{i}+(r_{0}-b_{i})e^{-a_{i}s}\big)\int_{-\infty}^{+\infty}f_{\mathcal{N}(m_{i,r_{0}}(\tau),\sigma_{i,r_{0}}^{2}(\tau))}(x)
R_{k,0,x}(s+h- \tau)dx d\tau \\
& +\sum_{k\in I}\int_{0}^{s} \frac{\dot{Q}_{ik}(\tau +u)}{1-H_{i}(u)}\int_{-\infty}^{+\infty}f_{\mathcal{N}(m_{i,r_{0}}(\tau ),\sigma_{i,r_{0}}^{2}(\tau))}(x)\Xi_{k,0,x}(s-\tau, s+h-\tau)dx d\tau
\end{aligned}
\end{equation}
where $\rho_{i,r_{0}}(s,s+h)$ is given by
\begin{equation}
\rho_{i,r_{0}}(s,s+h)= \Exp [r_{i,r_{0}}(s)r_{i,r_{0}}(s+h)] = \sigma_{i}^{2}e^{-a_{i}h} ,
\end{equation}
and $f_{\mathcal{N}(m_{i,r_{t}}(\tau),\sigma_{i,r_{t}}^{2}(\tau))}$ is the distribution function of a normal with mean $m_{i,r_{t}}(\tau)$ and variance $\sigma_{i,r_{t}}^{2}(\tau)$ (cf. eq. (\ref{Eq:meanVas}) and (\ref{Eq:varVas}) respectively) .
\end{Corollary}
We conclude this subsection with a Monte Carlo algorithm able to generate the trajectories of the Vasicek model modulated by a semi-Markov process of kernel $\mathbf{Q}$ in the time interval $[0, T]$. The algorithm consists in repeated random sampling to compute successive visited states of the random variables $\{J_{0}, J_{1},...\}$, the jump times $\{T_{0},T_{1},...\}$ and the force of interest process $r(t)$ up to the time $T$.\\
The algorithm consists of 5 steps:
\begin{enumerate}
\item Set $n=0$, $J_{0}=i$, $T_{0}=0$, $r(0)=r_{0}$, horizon time$=T$; discretization step$=h$
\item Sample $J$ from $p_{J_{n},\cdot}$ and set $J_{n+1}=J(\omega)$;
\item Sample $W$ from $G_{J_{n},J_{n+1}}(\cdot)$ and set $T_{n+1}=T_{n}+W(\omega)$;
\item For each $i=T_{n}:h:\max(T,T_{k+1}-1)$\\
$\quad$  \,\,\, Sample $N$ from $\mathcal{N}(0,h)$ \\
$\quad$  \,\,\, Set $r(i+1)=b_{J_{n}}+e^{-a_{J_{n}}}(r(i)-b_{J_{n}})+e^{-a_{J_{n}}}\sigma_{J_{n}}N(i+1)$
\item if $T_{n+1}\geq T$ stop\\
$\quad$  \,\,\, else Set $n=n+1$ and go to 2).
\end{enumerate}

\subsection{Hull and White modulated model}
Let's now assume that the process $r_{i,r_{0}}(t)$ satisfies
\begin{displaymath}
\left\{ \begin{array}{l}
dr_{i,r_{0}}(t) = (\alpha_{i}(t) - \beta_{i}(t)r_{i,r_{0}}(t))dt + \sigma_{i}(t) dW_{t}\\
r_{i,r_{0}}(0) = r_{0}
\end{array} \right.
\end{displaymath}
where $\alpha_{i}$, $\beta_{i}$ and $\sigma_{i}$ are deterministic function of time. The solution of this stochastic differential equation can be expressed (see \cite{Shreve}) as
\begin{eqnarray}
r_{i,r_{0}}(t) = e^{-k_{i}(t)} \left[ r_{0} + \int_{0}^{t}e^{k_{i}(u)}\alpha_{i}(u)du + \int_{0}^{t}e^{k_{i}(u)}\sigma_{i}(u)dW(u) \right] ,
\end{eqnarray}
where
\begin{eqnarray}
k_{i}(t) = \int_{0}^{t} \beta_{i}(u) du .
\end{eqnarray}
Note that $(r_{i,r_{0}}(t))_{t \in [0,T]}$ is a Gaussian process whose mean is given by
\begin{eqnarray}
m_{i,r_{0}}(t) = e^{-k_{i}(t)} \left[ r_{0} + \int_{0}^{t}e^{k_{i}(u)}\alpha_{i}(u)du \right] ,
\end{eqnarray}
and its variance is given by
\begin{eqnarray}
\sigma^{2}_{i,r_{0}}(t) = e^{-2k_{i}(t)}  \int_{0}^{t}e^{2k_{i}(u)}\sigma_{i}(u)^{2}du .
\end{eqnarray}
Morover the process $(\int_{0}^{t}r_{i,r_{0}}(s)ds)_{t \in [0,T]}$ is Gaussian as well, with mean
\begin{eqnarray}
\Exp \left[ \int_{0}^{t}r_{i,r_{0}}(s)ds \right] = \int_{0}^{t} e^{-k_{i}(t)} \left[ r_{0} + \int_{0}^{s}e^{k_{i}(u)}\alpha_{i}(u)du \right] ds ,
\end{eqnarray}
and variance
\begin{eqnarray}
Var \left( \int_{0}^{t}r_{i,r_{0}}(s)ds \right) = \int_{0}^{t} e^{2k_{i}(u)} \sigma_{i}^{2}(u) \left( \int_{0}^{t}e^{-k_{i}(s)}ds\right)^{2} du .
\end{eqnarray}
Then, we are able to express the $n^{th}$ moment of a ZCB value between two consecutive renewal moment in an analytical form
\begin{eqnarray}
& & B^{(n)}_{i,r_{0}}(0,t) = \Exp \left[ exp \{-n \int^{T}_{0} r_{i,r_{0}}(t)dt \} \right] \nonumber \\
&=& exp \left\{ - \Exp \left[ \int^{T}_{0} r_{i,r_{0}}(t)dt \right] + \frac{1}{2}  Var \left[ \int_{0}^{T} r_{i,r_{0}}(t) dt \right] \right\} .
\end{eqnarray}
In analogy to the Vasicek model we could give similar results for the value of the $n^{th}$ moment of ZCB in an arbitrary time interval and the first two moments of the force of interest. We omit here the details for the sake of brevity.

\subsection{CIR modulated model}
In this model we assume that the dynamics of the force of interest is described, between two consecutive renewal moments, by the following stochastic differential equation
\begin{displaymath}
\left\{ \begin{array}{l}
dr_{i,r_{0}}(t) = (a_{i} - b_{i}r_{i,r_{0}}(t))dt + \sigma_{i} \sqrt{r_{i,r_{0}}(t)} dW_{t}\\
r_{i,r_{0}}(0) = r_{0}
\end{array} \right.
\end{displaymath}
here, $a_{i}$ and $\sigma_{i}$ are non negative constants while $b_{i} \in \Real$.  This stochastic differential equation, for any $r_{0} \in \Real$, has a unique solution (see \cite{Ikeda}). We will not write the explicit solution of this equation, but following the approach of \cite{Lamberton}, we will study the property of the distribution of $r_{i,r_{0}}(t)$. To this aim, let first study the distribution of $(r_{i,r_{0}}(t),\int_{0}^{t}r_{i,r_{0}}(s)ds)$. It is possible to show (see \cite{Lamberton}) that
\begin{eqnarray}
\Exp \left( e^{-\lambda r_{i,r_{0}}(t)} e^{- \mu \int_{0}^{t}r_{i,r_{0}}(s)ds} \right) = \exp(-a_{i}\phi_{i,\lambda,\mu}(t)) \exp(-r_{0}\psi_{i,\lambda,\mu}(t))
\end{eqnarray}
where the functions $\phi_{i,\lambda,\mu}(t)$ and $\psi_{i,\lambda,\mu}(t)$ are given by
\begin{eqnarray}
\phi_{i,\lambda,\mu}(t) = - \frac{2}{\sigma_{i}^{2}}\log \left( \frac{2\gamma_{i} e^{\frac{t(\gamma_{i} + b_{i})}{2}}}{\sigma_{i}^{2}\lambda (e^{\gamma_{i} t} - 1) + \gamma_{i} - b_{i} + e^{\gamma_{i} t} ( \gamma_{i} + b_{i})} \right)
\end{eqnarray}
and
\begin{eqnarray}
\psi_{i,\lambda,\mu}(t) = \frac{\lambda (\gamma_{i} + b_{i} + e^{\gamma_{i} t}(\gamma_{i} - b_{i})) + 2\mu (e^{\gamma_{i} t} - 1)}{\sigma_{i}^{2}\lambda (e^{\gamma_{i} t} - 1) + \gamma_{i} - b_{i} + e^{\gamma_{i} t} ( \gamma_{i} + b_{i})}
\end{eqnarray}
with $\gamma_{i} = \sqrt{b_{i}^{2} + 2 \sigma_{i}^{2}\mu}$ .  Using this result we can explicitly evaluate the Laplace transform of  $r_{i,r_{0}}(t)$, indeed by putting $\mu = 0$ in the above expression we get 
\begin{eqnarray}
\Exp \left( e^{-\lambda r_{i,r_{0}}(t)} \right) = \left( \frac{2b_{i}}{\sigma_{i}^{2}\lambda (1 - e^{-b_{i}t}) + 2b_{i}} \right)^{\frac{2a_{i}}{\sigma_{i}^{2}}} \exp \left( -r_{0} \frac{2 \lambda b_{i} e^{-b_{i}t}}{\sigma_{i}^{2}\lambda (1 - e^{-b_{i}t}) + 2b_{i}} \right) .
\end{eqnarray}
Moreover, we can obtain an analytic representation of the $n^{th}$ moment of a ZCB value between two renewal moments, in fact setting $\lambda = 0$ and $\mu = n$ we have
\begin{eqnarray}
& & B_{i,r_{0}}^{(n)}(0,t) = \Exp \left( e^{- n \int_{0}^{t}r_{i,r_{0}}(s)ds} \right) \nonumber\\
& =& \left( \frac{2\gamma_{i} e^{\frac{t(\gamma_{i} + b_{i})}{2}}}{\gamma_{i} - b_{i} + e^{\gamma_{i} t} ( \gamma_{i} + b_{i})} \right)^{\frac{2a_{i}}{\sigma_{i}^{2}}} \exp \left( -r_{0} \frac{2 (e^{\gamma_{i} t} - 1)}{\gamma_{i} - b_{i} + e^{\gamma_{i} t} ( \gamma_{i} + b_{i})} \right) ,
\end{eqnarray}
where $\gamma_{i} = \sqrt{b_{i}^{2} + 2 \sigma_{i}^{2}n}$ . Starting from here we can obtain similar result as for the Vasicek model for the $n^{th}$ moment of a ZCB in an arbitrary time interval and the first two moments of the force of interest. We omit here the details for the sake of brevity.


\bibliographystyle{model2-names}

\end{document}